# Integrated lithium niobate photonic computing circuit based on efficient and high-speed electro-optic conversion

Yaowen Hu[1,2,\*,#], Yunxiang Song[1,3,\*], Xinrui Zhu[1], Xiangwen Guo[4], Shengyuan Lu[1], Qihang Zhang[5], Lingyan He[6], C. A. A. Franken[1,7], Keith Powell[1], Hana Warner[1], Daniel Assumpcao[1], Dylan Renaud[1], Ying Wang[6], Letícia Magalhães[1], Victoria Rosborough[8], Amirhassan Shams-Ansari[1], Xudong Li[1], Rebecca Cheng[1], Kevin Luke[8], Kiyoul Yang[1], George Barbastathis[9], Mian Zhang[6], Di Zhu[10,11], Leif Johansson[8], Andreas Beling[4], Neil Sinclair[1], Marko Loncar[1, #]

[1]*John A. Paulson School of Engineering and Applied Sciences, Harvard University, Cambridge, MA 02138, USA*
[2]*State Key Laboratory for Mesoscopic Physics and Frontiers Science Center for Nano-optoelectronics, School of Physics, Peking University, Beijing 100871, China*
[3]*Quantum Science and Engineering, Harvard University, Cambridge, MA 02138, USA*
[4]*Department of Electrical and Computing Engineering, University of Virginia, Charlottesville, Virginia 22903, USA*
[5]*Department of Electrical Engineering and Computer Science, Massachusetts Institute of Technology, Cambridge, MA 02139, USA*
[6]*HyperLight Corporation, 501 Massachusetts Ave, Cambridge, MA 02139, USA*
[7]*Laser Physics and Nonlinear Optics Group, Department of Science and Technology, MESA+ Institute of Nanotechnology, University of Twente, Enschede, The Netherlands*
[8]*Freedom Photonics, 41 Aero Camino, Goleta, CA, USA*
[9]*Department of Mechanical Engineering, Massachusetts Institute of Technology, Cambridge, MA 02139, USA*
[10]*Department of Materials Science and Engineering, National University of Singapore, Singapore 117575, Singapore*
[11]*Institute of Materials Research and Engineering (IMRE), Agency for Science, Technology and Research (A\*STAR), Singapore 138634, Singapore*
[\*]*These authors contributed equally.* [#]*Corresponding authors: yaowenhu@pku.edu.cn; loncar@seas.harvard.edu*

**The surge in artificial intelligence applications, encompassing machine vision, autonomous driving, remote sensing, and process simulation, relies on efficient and large-scale data processing. This requires highly scalable computation with high speed and low energy consumption, calling for innovative solutions to overcome the limitations of electronic systems[1–9]. Integrated photonics may play a pivotal role in addressing this challenge, leveraging the intrinsic properties of photons, such as natural parallelism, high bandwidth, and low latency[10–15]. While current photonic computing approaches show promise, they are critically limited by the speed and energy consumption of mapping information from the electronic to optical domain (i.e., electro-optic conversion)[1–3,6–8,16]. Here we show a photonic computing accelerator utilizing a system-level thin-film lithium niobate circuit which overcomes this limitation. Leveraging the strong electro-optic (Pockels) effect and the scalability of this platform, we demonstrate photonic computation at speeds up to 1.36 TOPS while consuming 0.057 pJ/OP. Our system features more than 100 thin-film lithium niobate high-performance components working synergistically, surpassing state-of-the-art systems on this platform. We further demonstrate binary-classification, handwritten-digit classification, and image classification with remarkable accuracy, showcasing our system's capability of executing real algorithms. Finally, we investigate the opportunities offered by combining our system with a hybrid-integrated distributed feedback laser source and a heterogeneous-integrated modified uni-traveling carrier photodiode. Our results illustrate the promise of thin-film lithium niobate as a computational platform, addressing current bottlenecks in both electronic and photonic computation. Its unique properties of high-performance electro-optic weight encoding and conversion, wafer-scale scalability, and compatibility with integrated lasers and detectors, position thin-film lithium niobate photonics as a valuable complement to silicon photonics, with extensions to applications in ultrafast and power-efficient signal processing[17] and ranging[18,19].**

# Introduction

The desire for intelligent systems capable of autonomous learning, reasoning, and adaptation has fueled significant advancements in artificial intelligence (AI), transforming various application landscapes. As the demand for extensive computational resources rapidly grows, traditional electronic computing approaches for AI are reaching their inherent limits in speed and energy efficiency for parallel processing. This limitation has stimulated the exploration of novel computing architectures employing advanced computational paradigms. One example is photonic computing[10–12,20], which seeks to harness the unique properties of photons, such as high bandwidth enabled by the high optical carrier frequency and inherent parallelism that leverages the frequency and polarization degrees of freedom, to perform computational tasks traditionally executed by electronic systems, thereby achieving unprecedented speeds and energy efficiencies.

Driven by these unique advantages, as well as the rapid development of integrated photonics, photonic computing has emerged as a promising solution for realizing next-generation computing accelerators. Demonstrations using Mach-Zehner interferometer arrays[1,21–23], free space optics[6,7,16,24,25], silicon photonics with on-chip attenuators, photodetectors, and ring banks[8,26–28], VCSELs with free space optics[29], as well as parallel processors utilizing frequency multiplexing[2,3,30,31] and phase change materials[3,32,33], showcase the versatility and potential of the photonics approach. Despite these advances, a critical challenge persists. Low speed and high energy consumption are associated with the electro-optic (EO) conversion of data from electronic memory into the optical domain. However, such EO conversion is unavoidable, as the majority of data is stored and processed electronically. Current approaches utilizing variable attenuators or fast electro-optic modulators[1–3,3,21–23,26–30,32–34], typically implemented in silicon photonics or bulk systems, are capable of preparing data at rates ranging from kHz to GHz but are plagued by high electronic energy consumption and optical loss. Alternatively, approaches using static amplitude masks or spatial light modulators can be more energy efficient yet suffer from slow refresh rates[6–8,16,24,25]. Therefore, prior works only demonstrated either high speed or low energy consumption, but not both simultaneously. As a result, EO conversion remains the most crucial hurdle for practical implementations of photonic computing in real-world applications. Ultimately, high-performance EO modulation is necessary to achieve a high optical data rate that can match the low optical latency while maintaining a low energy budget. However, a photonic accelerator that combines high-speed processing with low energy consumption, addressing the crucial challenge of EO conversion, is still missing.

Here we demonstrate high-speed and energy-efficient photonic computation, using the Pockels (EO) effect in a thin-film lithium niobate (TFLN) circuit. Recent progress in the TFLN photonic platform has enabled development of groundbreaking EO devices including modulators[35–37], frequency shifters[38], and comb generators[39,40], among others. The remarkable capabilities of these devices stem from the strong EO effect and are facilitated by the tight confinement of both optical and electronic modes as well as low propagation loss. By seamlessly integrating these high-performance EO devices into an optimized large-scale photonic circuit, we show a scalable TFLN computing accelerator for matrix-vector-multiplication (MVM) that addresses the current speed and power limitations of state-of-the-art computing architectures.

## Results

### Accelerator architecture

The accelerator imprints electronic data onto the optical domain by modulating optical amplitudes, performs computations of the data through successive electro-optic modulation, and returns the result back to the electrical domain through optical-to-electronic conversion (Fig. 1a). This workflow is implemented using two high-performance TFLN amplitude modulators connected in series, and photonic advantage is achieved through ultra-efficient and massively parallel multiplexing in the time domain. In the context of fully-connected deep neural networks (DNNs), where MVM for data inference is paramount, an input data vector may be encoded onto the amplitude of light by the first TFLN amplitude modulator. The data-modulated light is then fanned-out into distinct spatial channels, each channel subsequently mapping a machine-learned weight vector onto the data vector using a second TFLN amplitude modulator. The intensity of light passing through both modulators is thus proportional to the product of signals applied to each modulator, which amounts to multiplication of the weight vector with the data vector. This intensity is detected using a photodetector and summed electronically, converting the result back into the electrical domain.

### Photonic computing circuit implementation on TFLN

To demonstrate the proposed computing core, we developed individual photonic-integrated components and combined them into a TFLN circuit on a single chip. Our accelerator contains $M = 2$ computing cores, each with $N = 16$ channels realized by a spatial fan-out. A total of $M \times N = 32$ parallel channels for multiplication are thus established owing to the large set of synchronously-functional EO modulators (32 modulators) with 32 on-chip terminators. Further, numerous passive optical elements (e.g. 34 grating couplers, and 30 Y splitters for fan-out splitter trees) are fabricated through high-quality, direct dry etching of TFLN, and smooth waveguide sidewalls over the entire circuit lead to a low propagation optical loss of 0.28 dB/cm, achieved by optimizing the reactive ion etching process (Fig. 2d). Low propagation loss critically enables a large-scale TFLN circuit such as ours, as well as energy-efficient computing through one-to-sixteen fan-out splitter trees with added insertion loss of 0.135 dB per channel. The microwave circuit for modulation is accomplished using two layers of gold (labeled with $h_1$ and $h_2$ in Fig. 2e) with nickel-chromium (NiCr) resistors (Fig. 2e). The bottom layer of gold is required for efficient modulation and direct-current (DC) tunability, while the top layer interfaces with resistors. The use of NiCr ensures high-resistance terminators with high-power handling ability and minimal feature size (Fig. 2f). Each 1-cm long modulator features >40 GHz bandwidth (limited by the 40-GHz detector used), >20 dB microwave reflection suppression, $V_\pi \cdot L \sim 2.2$ V·cm , and >20 dB extinction ratio (Fig. 2g, h).

### High-speed and energy-efficient photonic computing on TFLN

We first demonstrate high-speed photonic computing with low energy consumption (Fig. 3a). For generality, we generate random vectors of length 1,000 as data and weight vectors. Both vectors are encoded into the time domain at a rate $r = 43.8$ GOPS per channel and multiplied after light passes through the data and weight modulators. The measurements are in excellent agreement with expectation, verifying the high computational speed per channel of our circuit (Fig. 3c and see Supplementary Materials for details). Next, we evaluate the total circuit performance in this way and find that the total computational speed of our chip can be inferred as $31 \times r = 1.36$ TOPS, where only one of all 32 channels was not optimally functioning. Operating at such a high speed (43.8 GOPS per channel), we then evaluate the energy consumption while simultaneously assessing computational accuracy under varying optical power conditions. Our experiments reveal a minimal energy consumption of 0.057 pJ/OP, demonstrating robust computational accuracy with high energy efficiency for the entire photonic system (Fig. 3d and see "Computational accuracy" in Supplementary Materials for details). This assessment includes energy consumption of the pump laser, microwave energy dissipated by modulators and detectors, as well as energy consumption of digital-to-analog and analog-to-digital converters (see "System characterization: speed and energy consumption calculations" in Supplementary Materials for details). Altogether, our results confirm the high-speed and energy-efficient multiplication of generalized vectors using the TFLN computing system.

**Inference tasks**

We utilize the TFLN circuit to perform computations within real algorithms. First, we tackle a simple binary classification problem (Fig. 4) over points lying in a two-dimensional plane labeled by "exclusive-or" rule. One such data point $\vec{x} = \begin{pmatrix} x_1 \\ x_2 \end{pmatrix}$ is input into the circuit for a small number of computing operations to be carried out, and nonlinear activation via an electronic computer is used to obtain the label. The sign of the label indicates the category of the data point inferred by photonics (Fig. 4a). A total of 400 data points is tested, with the classification results shown in the 2D plane (Fig. 4b), achieving an overall accuracy of 93.8% (93.5% on the electronic computer). Further analysis histograms reveal an accuracy of 98.6% (99.1% on the electronic computer) for data points with positive ground truth and 88.4% (87.3% on the electronic computer) for data points with negative ground truth (Fig. 4c). The photonic computing system is thus effective for performing binary classification tasks, supported by competitive classification accuracies compared to those achieved by traditional electronic computers.

Next, we apply the photonic accelerator to a handwritten digit classification problem over the Modified National Institute of Standards and Technology (MNIST) dataset. A two-layer feedforward model is chosen to perform the classification (Fig. 5a), with input vectors of size 784 temporally encoded in in layer 1, and 10 neurons in layer 2. These input vectors undergo photonic MVM processing followed by electronic softmax activations. Photonic and electronic confusion matrices (Fig. 5b) generated over a 500-image test set indicate that our circuit can efficiently perform computing operations for machine learning inference, as the circuit achieves a classification accuracy of 88%, compared to 92% obtained from electronic computing. In addition to binary classification, our results demonstrate the versatility of our photonic computing system in handling various models, including neural network tasks with vectors of arbitrarily long length. Importantly, we also evaluated the system stability by running identical computation tasks on the photonic circuit for over 20 hours, and we observed excellent stability with a standard deviation in the error fluctuation of 0.04% (Fig. 5c).

Finally, to assess the practical utility of our computing system in solving real-world problems, we extend our demonstrations to image classification over the Canadian Institute for Advanced Research 10-class (CIFAR-10) dataset. Images undergo preprocessing with a convolution layer, a step that could be directly mapped to MVMs or accelerated on TFLN by specializing the circuit for convolution operations[2,3,30]. In our setup, the circuit is used to propagate preprocessed images through the fully connected layers of the convolutional neural network (CNN). The first layer consists of an input vector of size 1024 encoded in time, which is subsequently transformed into an intermediate layer with 128 neurons, and then a final output layer with 10 neurons representative of the ten classes. We tested images in all 10 categories. Examples including trucks, cats, birds, automobiles, frogs, deer, ships, airplanes, and horses are presented, with accurate classification results (Fig. 5d). This confirms our system's potential for complex image recognition and analysis typically requiring large, multi-layer neural networks since the correct classification of one image requires a significant amount of computing operations.

**Hybrid- and Heterogenous- integrated TFLN photonic computing circuit**

Next generation high-performance and low power consumption photonic computing circuits would feature lasers and detectors integrated on TFLN chips, alongside EO and passive TFLN components. Importantly, wafer-scale fabrication of TFLN (Fig. 6a) is ideally suited for the batch production of stand-alone optical computing cores (Fig. 6b, c), and it would unlock significant potential for massive spatial parallelism in conjunction with temporal-multiplexing and ultra-efficient EO conversion, characteristics unique to the TFLN photonic computing approach demonstrated here.

As an important first step in this direction, we fabricated a TFLN photonic computing accelerator chip with heightened degree of integration featuring a single spatial channel. Our chip consists of previously mentioned active and passive TFLN components, with the addition of bilayer-taper edge couplers that are used for hybrid-integration of a distributed feedback (DFB) laser source with the TFLN chip, and heterogeneous-integrated modified uni-traveling-carrier photodiodes (MUTC-PDs) evanescently coupled to TFLN signal waveguides (Fig. 6d-h). We calibrate the DFB laser power against injection current and find a lasing threshold of about 50 mA (Fig. 6i) with 0.25 W/A slope efficiency. We also measure the MUTC-PD dark current as function of bias voltage (Fig. 6j) and find a low dark current of 2.7 nA when reverse biased at -2 V for high-speed operation (see Supplementary Materials for further details on DFB characterization and PD fabrication). Given these performances, we evaluate key computing performance metrics, such as high-speed and low energy consumption computations, by operating single spatial channel at a rate $r$ =10.33 GOPS (Fig. 6k) and assessing the computational accuracy vs. various full-system energy consumptions (Fig. 6l). We show high-fidelity encoding and multiplication of pairs of random vectors, as well as computational accuracies characterized by $\sigma$ <0.5% across all energy consumptions attempted. Here, the reduction in demonstrated data rate (compared to previously demonstrated 43.8 GOPS per channel) is attributed to imperfect termination resistance. The larger energy consumptions overall (compared to previously demonstrated 0.057 pJ/OP) are mostly due to absence of spatial multiplexing in our proof-of-concept demonstration. We also note that the DFB source operated just above lasing threshold already provides significantly more optical power than needed for accurate computations over a single spatial channel.

**Conclusion and outlook**

In conclusion, our work presents a photonic computing accelerator on TFLN, capable of performing increasingly complex algorithmic tasks, from binary classification, handwritten digit classification, to actual image classification. Unlike accelerators designed for specific applications such as convolution or vision-based tasks, our accelerator is well suited for general computing tasks. Our demonstration is enabled by a large scale, system-level TFLN circuit, the first of its kind. Leveraging the TFLN platform, we critically include the process of electro-optic conversion in our demonstration to unambiguously achieve high speed and low energy consumption photonic computing for the first time, with a performance comparable to the state-of-the-art electronics. This is made possible by increasing the scale of TFLN integration while maintaining the high performance of individual components. We then significantly advance our system by fabricating on-chip detectors and replacing the bench-top laser with DFB source butt-coupled to our TFLN chip. We also demonstrated its potential for high-speed and low energy consumption operation. Future improvements to enhance performance metrics to match or exceed state-of-the-art photonic accelerators are straightforward: reduce the half-wave voltage ($V_\pi$) to 1V[36], increase the bandwidth to 100 GHz[36,41], and minimize the optical propagation loss of waveguides to 0.03 dB/cm [42,43], each of which has been previously demonstrated in TFLN devices. Transitioning the accelerator to visible wavelengths may further reduce energy consumption using sub-volt TFLN modulators[44,45], and utilizing the frequency degree-of-freedom with TFLN microcombs may extend applicability to photonic convolution acceleration. For a comprehensive contextualization against state-of-the-art electronic and photonic systems, as well as future projections in performance, see "System characterization: speed and energy consumption calculations" in Supplementary Materials.

It is worth noting that our approach to address the EO conversion challenge in photonic computing is compatible with other photonic accelerators. Thus, our work may motivate novel hybrid approaches (e.g. combining TFLN EO conversion with free space optics for computation[6,7,16,24,25]) that require ultrahigh bandwidth and low power modulation for data encoding. TFLN stands out as the most powerful platform for this task, owing to exceptional EO conversion capability, thus offering the potential to overcome existing speed and energy bottlenecks in various optical computation schemes. The full potential of the platform can be unlocked by leveraging photonic-electronic integration, especially integration with high-speed electronic circuits, including multi-channel DACs, ADCs, and FPGAs. We believe that TFLN-based photonic computing may hold great promise for applications in vision[6], sensing[9], ranging[18,19], and even quantum computing[46–48], and we hope that our work will stimulate further exploration of such applications.

Note: Photonic computing using TFLN platforms has also been demonstrated in Ref. [49–51]

## Acknowledgement

This work is supported by DARPA LUMOS (HR0011-20-C-0137), NSF (EEC-1941583, OMA-2137723, OMA-2138068), ONR (N00014-22-C-1041), NASA (231180A/ 80NSSC22K0262), Harvard Quantum Initiative, NRF and A*STAR Quantum Engineering Program (NRF2022-QEP2-01-P07), NRF Fellowship (NRF-NRFF15-2023-0005). These views, opinions and/or findings expressed are those of the authors and should not be interpreted as representing the official views or policies of the Department of Defense or the U.S. Government. Distribution Statement "A" (Approved for Public Release, Distribution Unlimited.)

## Competing interests

L.H., Y.W., K.L., M.Z., and M.L. are involved in developing lithium niobate technologies at HyperLight Corporation

## References

1. Shen, Y. *et al.* Deep learning with coherent nanophotonic circuits. *Nat. Photonics* **11**, 441–446 (2017).

2. Xu, X. *et al.* 11 TOPS photonic convolutional accelerator for optical neural networks. *Nature* **589**, 44–51 (2021).

3. Feldmann, J. *et al.* Parallel convolutional processing using an integrated photonic tensor core. *Nature* **589**, 52–58 (2021).

4. Wan, W. *et al.* A compute-in-memory chip based on resistive random-access memory. *Nature* **608**, 504–512 (2022).

5. Yao, P. *et al.* Fully hardware-implemented memristor convolutional neural network. *Nature* **577**, 641–646 (2020).

6. Chen, Y. *et al.* All-analog photoelectronic chip for high-speed vision tasks. *Nature* **623**, 48–57 (2023).

7. Lin, X. *et al.* All-optical machine learning using diffractive deep neural networks. *Science* **361**, 1004–1008 (2018).

8. Ashtiani, F., Geers, A. J. & Aflatouni, F. An on-chip photonic deep neural network for image classification. *Nature* **606**, 501–506 (2022).

9. Wright, L. G. *et al.* Deep physical neural networks trained with backpropagation. *Nature* **601**, 549–555 (2022).

10. Nahmias, M. A. *et al.* Photonic Multiply-Accumulate Operations for Neural Networks. *IEEE J. Sel. Top. Quantum Electron.* **26**, 1–18 (2020).

11. Wetzstein, G. *et al.* Inference in artificial intelligence with deep optics and photonics. *Nature* **588**, 39–47 (2020).

12. Shastri, B. J. *et al.* Photonics for artificial intelligence and neuromorphic computing. *Nat. Photonics* **15**, 102–114 (2021).

13. Sui, X., Wu, Q., Liu, J., Chen, Q. & Gu, G. A Review of Optical Neural Networks. *IEEE Access* **8**, 70773–70783 (2020).

14. McMahon, P. L. The physics of optical computing. *Nat. Rev. Phys.* **5**, 717–734 (2023).

15. Berggren, K. *et al.* Roadmap on emerging hardware and technology for machine learning. *Nanotechnology* **32**, 012002 (2021).

16. Zhou, T. *et al.* Large-scale neuromorphic optoelectronic computing with a reconfigurable diffractive processing unit. *Nat. Photonics* **15**, 367–373 (2021).

17. Marpaung, D., Yao, J. & Capmany, J. Integrated microwave photonics. *Nat. Photonics* **13**, 80–90 (2019).

18. Riemensberger, J. *et al.* Massively parallel coherent laser ranging using a soliton microcomb. *Nature* **581**, 164–170 (2020).

19. Zhang, X., Kwon, K., Henriksson, J., Luo, J. & Wu, M. C. A large-scale microelectromechanical-systems-based silicon photonics LiDAR. *Nature* **603**, 253–258 (2022).

20. Hamerly, R., Bernstein, L., Sludds, A., Soljačić, M. & Englund, D. Large-Scale Optical Neural Networks Based on Photoelectric Multiplication. *Phys. Rev. X* **9**, 021032 (2019).

21. Xu, S. *et al.* Optical coherent dot-product chip for sophisticated deep learning regression. *Light Sci. Appl.* **10**, 221 (2021).

22. Zhang, H. *et al.* An optical neural chip for implementing complex-valued neural network. *Nat. Commun.* **12**, 457 (2021).

23. Zhu, H. H. *et al.* Space-efficient optical computing with an integrated chip diffractive neural network. *Nat. Commun.* **13**, 1044 (2022).

24. Wang, T. *et al.* An optical neural network using less than 1 photon per multiplication. *Nat. Commun.* **13**, 123 (2022).

25. Wang, T. *et al.* Image sensing with multilayer nonlinear optical neural networks. *Nat. Photonics* **17**, 408 (2023).

26. Huang, C. *et al.* A silicon photonic–electronic neural network for fibre nonlinearity compensation. *Nat. Electron.* **4**, 837–844 (2021).

27. Tait, A. N. *et al.* Silicon Photonic Modulator Neuron. *Phys. Rev. Appl.* **11**, 064043 (2019).

28. Tait, A. N. *et al.* Neuromorphic photonic networks using silicon photonic weight banks. *Sci. Rep.* **7**, 7430 (2017).

29. Chen, Z. *et al.* Deep learning with coherent VCSEL neural networks. *Nat. Photonics* **17**, 723–730 (2023).

30. Bai, B. *et al.* Microcomb-based integrated photonic processing unit. *Nat. Commun.* **14**, 66 (2023).

31. Xu, X. *et al.* Photonic Perceptron Based on a Kerr Microcomb for High-Speed, Scalable, Optical Neural Networks. *Laser Photonics Rev.* **14**, 2000070 (2020).

32. Wu, C. *et al.* Programmable phase-change metasurfaces on waveguides for multimode photonic convolutional neural network. *Nat. Commun.* **12**, 96 (2021).

33. Feldmann, J., Youngblood, N., Wright, C. D., Bhaskaran, H. & Pernice, W. H. P. All-optical spiking neurosynaptic networks with self-learning capabilities. *Nature* **569**, 208–214 (2019).

34. Shi, B., Calabretta, N. & Stabile, R. Deep Neural Network Through an InP SOA-Based Photonic Integrated Cross-Connect. *IEEE J. Sel. Top. Quantum Electron.* **26**, 1–11 (2020).

35. Wang, C. *et al.* Integrated lithium niobate electro-optic modulators operating at CMOS-compatible voltages. *Nature* **562**, 101–104 (2018).

36. Xu, M. *et al.* Dual-polarization thin-film lithium niobate in-phase quadrature modulators for terabit-per-second transmission. *Optica* **9**, 61 (2022).

37. He, M. *et al.* High-performance hybrid silicon and lithium niobate Mach–Zehnder modulators for 100 Gbit s−1 and beyond. *Nat. Photonics* **13**, 359–364 (2019).

38. Hu, Y. *et al.* On-chip electro-optic frequency shifters and beam splitters. *Nature* **599**, 587–593 (2021).

39. Zhang, M. *et al.* Broadband electro-optic frequency comb generation in a lithium niobate microring resonator. *Nature* **568**, 373–377 (2019).

40. Hu, Y. *et al.* High-efficiency and broadband on-chip electro-optic frequency comb generators. *Nat. Photonics* **16**, 679–685 (2022).

41. Kharel, P., Reimer, C., Luke, K., He, L. & Zhang, M. Breaking voltage–bandwidth limits in integrated lithium niobate modulators using micro-structured electrodes. *Optica* **8**, 357 (2021).

42. Desiatov, B., Shams-Ansari, A., Zhang, M., Wang, C. & Lončar, M. Ultra-low-loss integrated visible photonics using thin-film lithium niobate. *Optica* **6**, 380 (2019).

43. Zhang, M., Wang, C., Cheng, R., Shams-Ansari, A. & Lončar, M. Monolithic ultra-high-Q lithium niobate microring resonator. *Optica* **4**, 1536 (2017).

44. Renaud, D. *et al.* Sub-1 Volt and high-bandwidth visible to near-infrared electro-optic modulators. *Nat. Commun.* **14**, 1496 (2023).

45. Xue, S. *et al.* Full-spectrum visible electro-optic modulator. *Optica* **10**, 125 (2023).

46. O'Brien, J. L., Furusawa, A. & Vučković, J. Photonic quantum technologies. *Nat. Photonics* **3**, 687–695 (2009).

47. Wang, J., Sciarrino, F., Laing, A. & Thompson, M. G. Integrated photonic quantum technologies. *Nat. Photonics* **14**, 273–284 (2020).

48. Kues, M. *et al.* Quantum optical microcombs. *Nat. Photonics* **13**, 170–179 (2019).

49. Lin, Z. *et al.* 120 GOPS Photonic Tensor Core in Thin-film Lithium Niobate for Inference and in-situ Training. Preprint at http://arxiv.org/abs/2311.16896 (2024).

50. Zheng, Y. *et al.* Photonic Neural Network Fabricated on Thin Film Lithium Niobate for High-Fidelity and Power-Efficient Matrix Computation. Preprint at http://arxiv.org/abs/2402.16513 (2024).

51. Chen, Z. *et al.* Hypermultiplexed Integrated-Photonics-based Tensor Optical Processor. Preprint at https://doi.org/10.21203/rs.3.rs-4778342/v1 (2024).

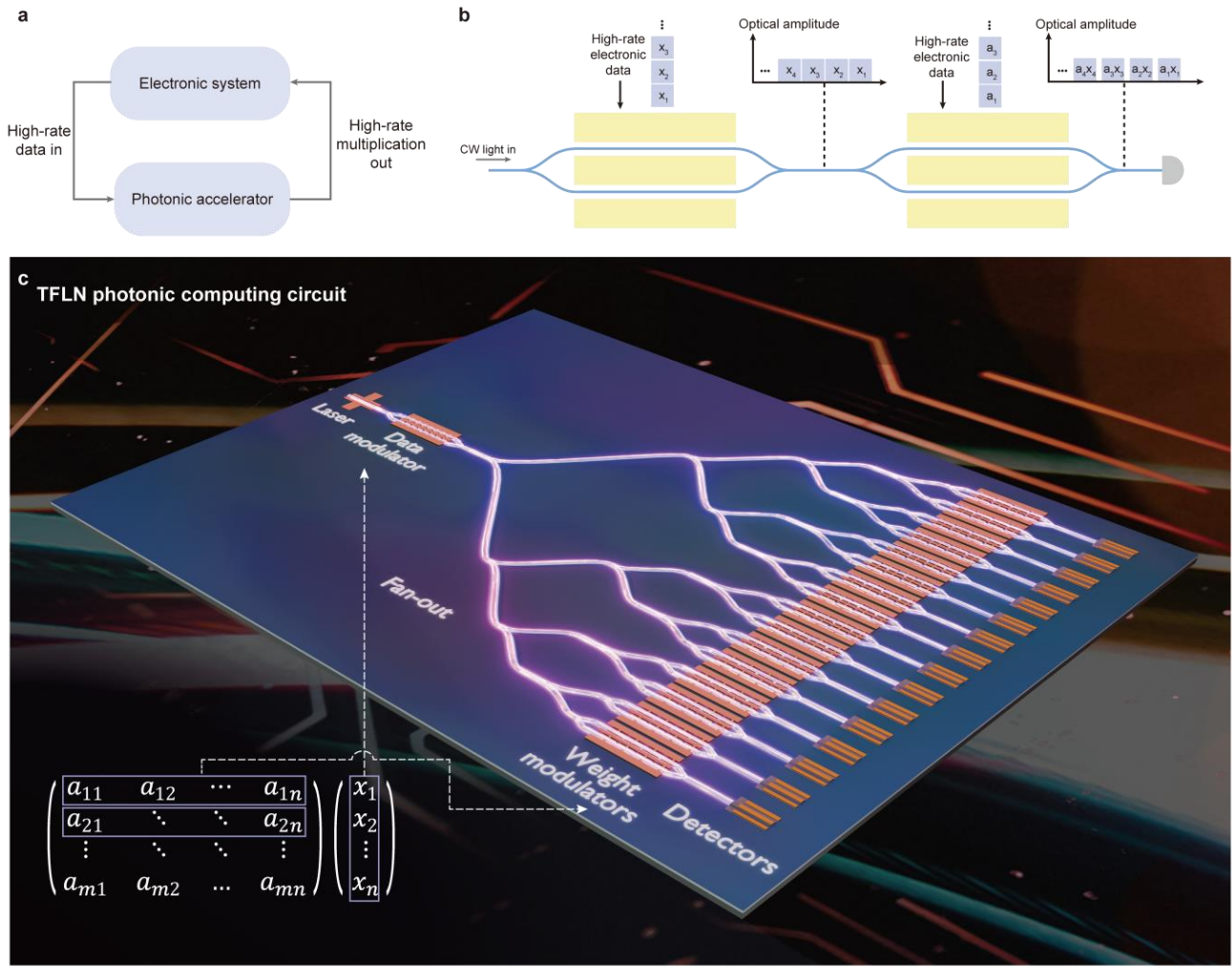

**Fig. 1 | Photonic computing accelerator on thin-film lithium niobate. a,** Concept of photonic accelerators. Data stored in the electronic system (e.g. a computer) are sent to the photonic accelerator at high rates and are converted into the optical domain. Parallel computations are then performed by the accelerator and results are returned to the electronic system. **b,** Illustration of the photonic computing working principle. Continuous-wave light passes through two cascaded amplitude modulators (AMs) which sequentially encode elements of $\vec{x}$ and $\vec{a}$ onto the amplitude of light, effectively performing element-wise multiplication of the two vectors. The components contributing to $\vec{x} \cdot \vec{a}$ are read out by optical-to-electronic conversion using a low-noise and high-speed detector, and electronic summation of these components finally yields $\vec{x} \cdot \vec{a}$. **c,** The vision for a fully integrated computing core based on TFLN photonics, consisting of laser, detectors, and TFLN modulators for high-speed and energy-efficient EO conversion and computation. An input vector is first encoded in the time domain of the optical field through an amplitude modulator and then fanned-out into $N$ spatial channels ($N = 16$ in this figure) to leverage massive spatial parallelism. In each channel, another amplitude modulator is used to multiply weights with the input vector. Finally, detectors convert the multiplication results back into electronic signals. In this work, the accelerator has $M = 2$ cores with $N = 16$ channels per core ($M = 2, N = 16$).

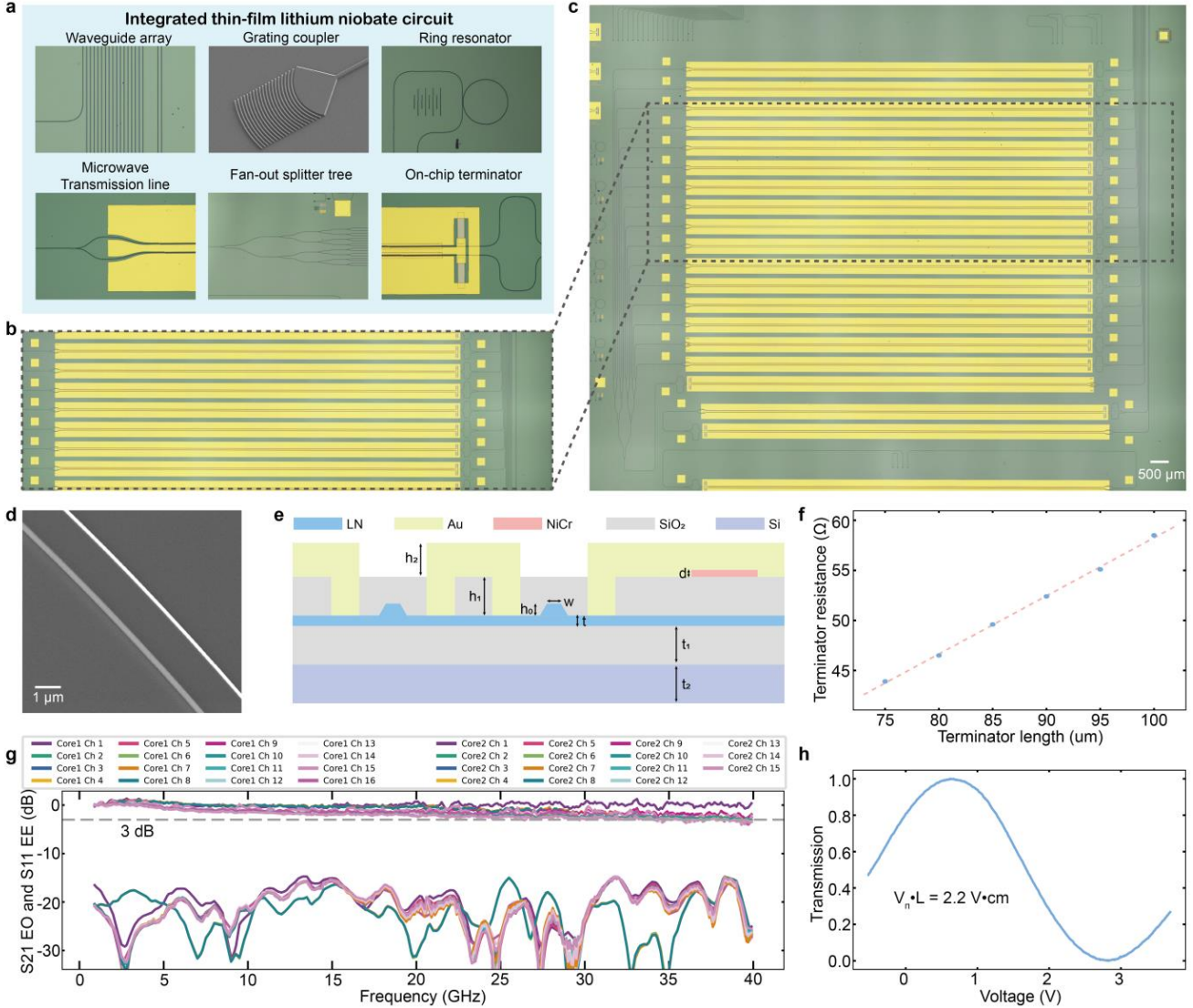

**Fig. 2 | Integrated thin-film lithium niobate photonic circuit implementation of computing core. a,** Optical microscope and scanning electron microscope images of the building blocks used in the integrated TFLN circuit: waveguide array (top left) for signal routing; grating coupler (top center) for efficient in- and out-coupling of light; ring resonator (top right) for evaluating the propagation loss and etch quality; fan-out splitter tree (bottom center) to distribute light into distinct spatial channels; microwave transmission line (bottom left) for delivering efficient electro-optic modulation; and on-chip terminator (bottom right) for high-quality microwave impedance matching. **b,** Optical microscope image of seven weight modulators in the TFLN accelerator. **c,** Full image of one computing core (our circuit contains two such computing cores on the same chip). **d,** Scanning electron microscope image of a high-quality waveguide featuring low propagation loss, enabling low optical energy consumption for the entire circuit. **e,** Cross-section illustration of the TFLN circuit, including gold, nickel-chromium (NiCr), lithium niobate, silicon dioxide, and silicon. $d = 100$ nm, $w = 1.5$ μm, $h_0 = 300$ nm, $h_1 = 1.0$ $\mu m$, $h_2 = 800$ nm, $t = 300$ nm, $t_1 = 4.7$ μm, $t_2 = 525$ μm. **f,** Measured terminator resistance vs. length. **g,** Electro-optic forward transmission ($S_{21}$ EO) and electric-electric input reflection ($S_{11}$ EE) response of modulators in our circuit. Our circuit features a combined total of 31 spatial channels (32 designed, but one failed during fabrication process). All modulators have a bandwidth beyond 40 GHz (measurement limited by our detector bandwidth). **h,** Representative $V_\pi \cdot L$ of modulators (length 1 cm) in our circuit.

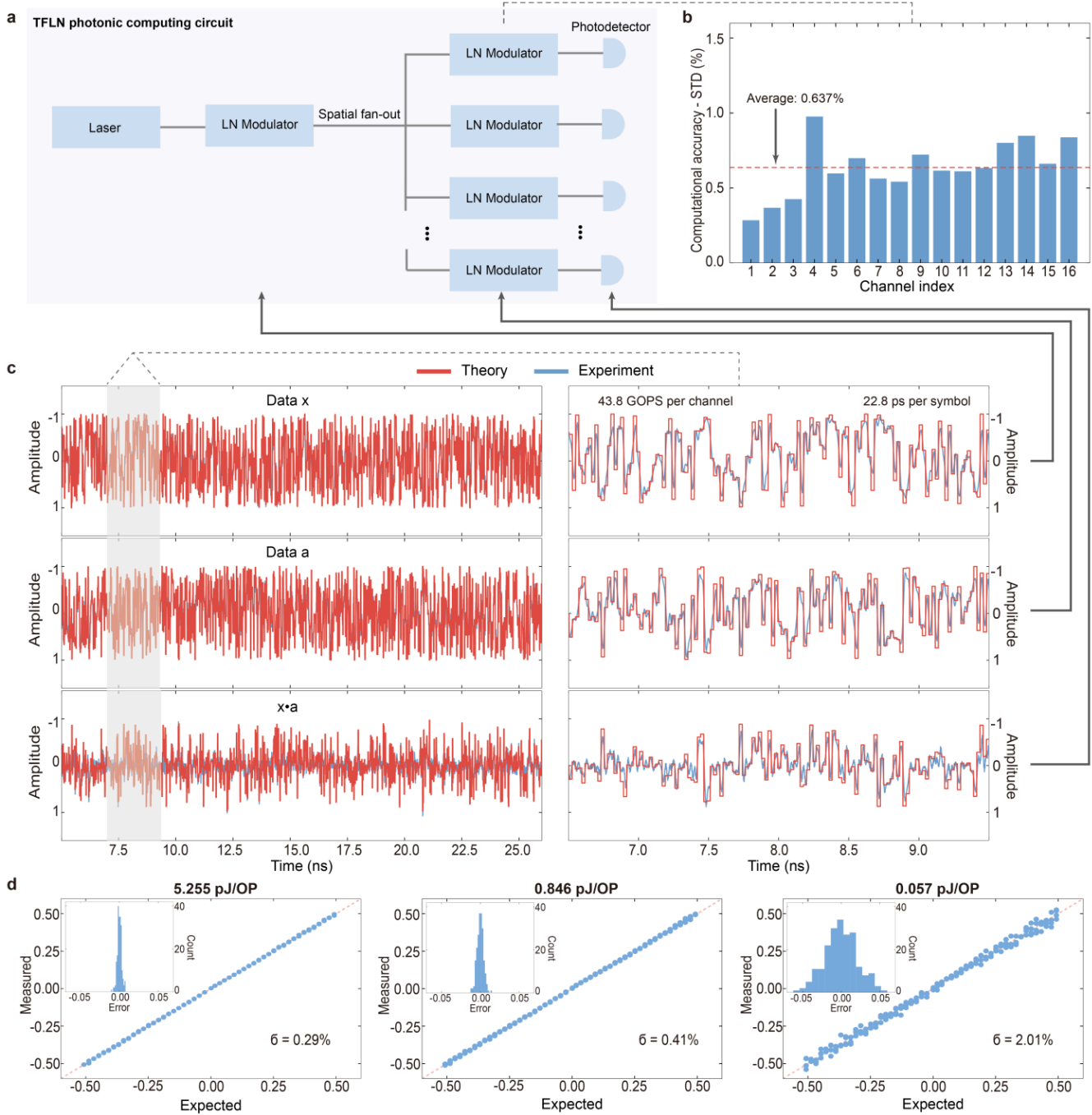

**Fig. 3 | High-speed and energy-efficient photonic computing on thin-film lithium niobate. a,** Two-dimensional illustration of the photonic computing core structure. **b,** Computational accuracy for different channels in one computing core. The differences in accuracy between channels is minimal though can be further reduced through fine tuning of the system operational parameters. **c,** Example waveforms of a temporally-multiplexed computing operation between two random vectors $\vec{x}$ and $\vec{a}$ with 22.8 ps/symbol. **d,** Computational accuracy vs. energy consumption by varying the optical power, showing a lowest energy consumption of 0.057 pJ/OP at 22.8 ps/symbol, while still maintaining excellent computational accuracy. The inset gives error of the computation (Error=Measured−Expected). The $\sigma$ is the standard deviation of the error.

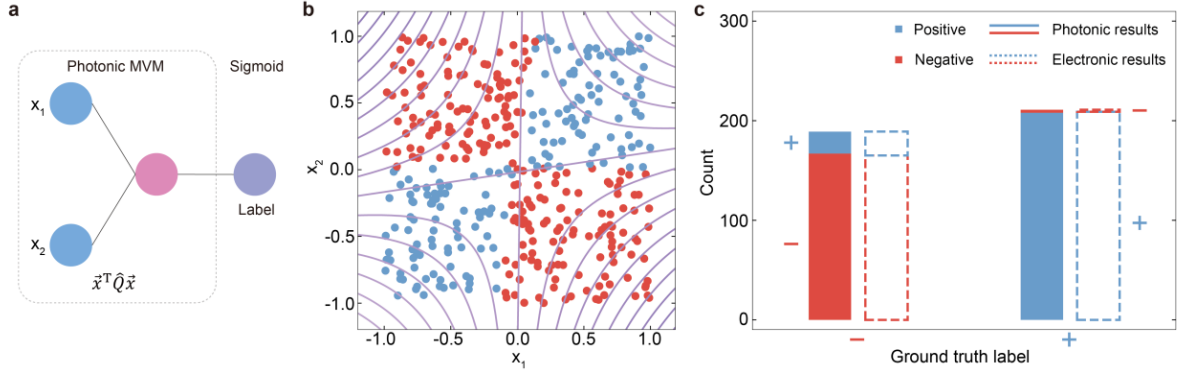

**Fig. 4 | Photonic binary classification. a,** Illustration of binary classification conducted over a data set consisting of two-dimensional vectors $\vec{x} = [x_1, x_2]^T$, where each vector is labeled positive or negative based on the exclusive-or condition. To infer the label of some $\vec{x}$, a number of computing operations are carried out by our photonic accelerator to compute $\vec{x}^T Q \vec{x}$ followed by electronic nonlinear activation. Here, $Q$ is a pre-trained kernel matrix and nonlinearity is electronically applied. Accuracy of inference thus depends on the accuracy of photonic computations. **b,** Classification of 400 randomly selected $\vec{x}$ in the problem space, colored by their photonics-inferred positive (blue) and negative (red) labels. **c,** Comparison between photonic and electronic classification results over the 400-vector test set, which shows excellent agreement between photonic and electronic computing. The photonic circuit (electronic computer) achieves a classification accuracy of 93.8% (93.5%).

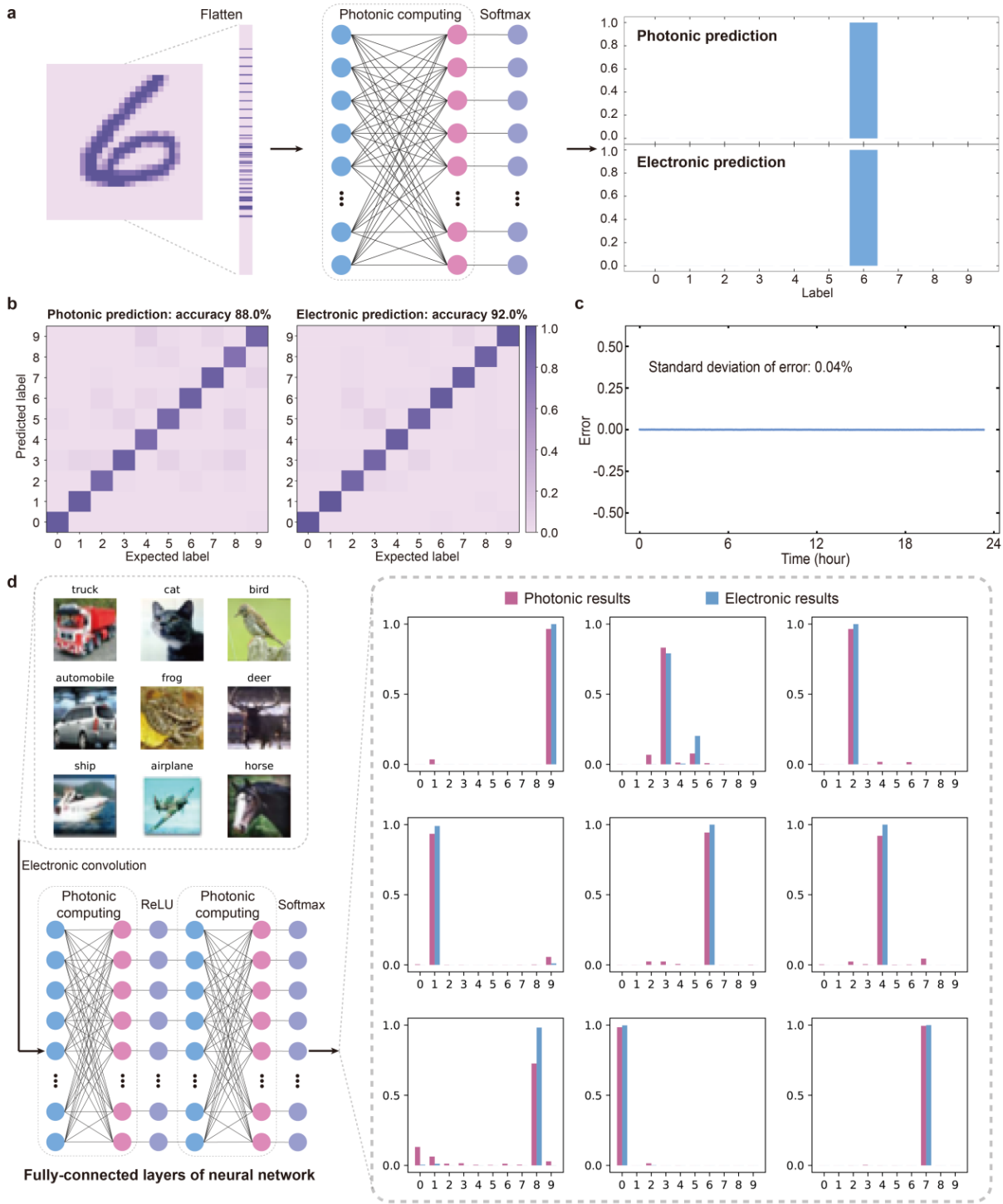

**Fig. 5 | Photonic computing for fully-connected layers of photonic neural networks. a,** Classification of an MNIST handwritten digit. The image is flattened into a single vector encoded in the time domain. An example image (number six) is shown on the left. A two-layer photonic neural network is used to perform the classification task (center). The photonic computing result at the end of the network is then sent back to the computer to perform a nonlinear activation (right). The final classification results agree well with the electronically computed result. **b,** Statistics of MNIST handwritten digit recognition. 500 MNIST images are selected as the test set and processed. The confusion matrices show an excellent classification accuracy of 88% using our circuit (92% using electronic computer). **c,** Stability. The computing core is programmed to continuously run

identical computing tasks over 20 hours, experiencing minimal fluctuations of 0.04%. **d,** Real image classification. Images resembling real-life objects are selected from the CIFAR-10 database and classified using a convolutional neural network. The images are preprocessed through convolution layers, flattened into vectors, and then sent into our circuit to be classified by the remaining fully-connected layers (bottom left). Nine example figures including a truck, cat, bird, automobile, frog, deer, ship, airplane, and horse (top left), together with the classification results are shown (right), indicating our photonic computing circuit can accurately compute large, multi-layer networks.

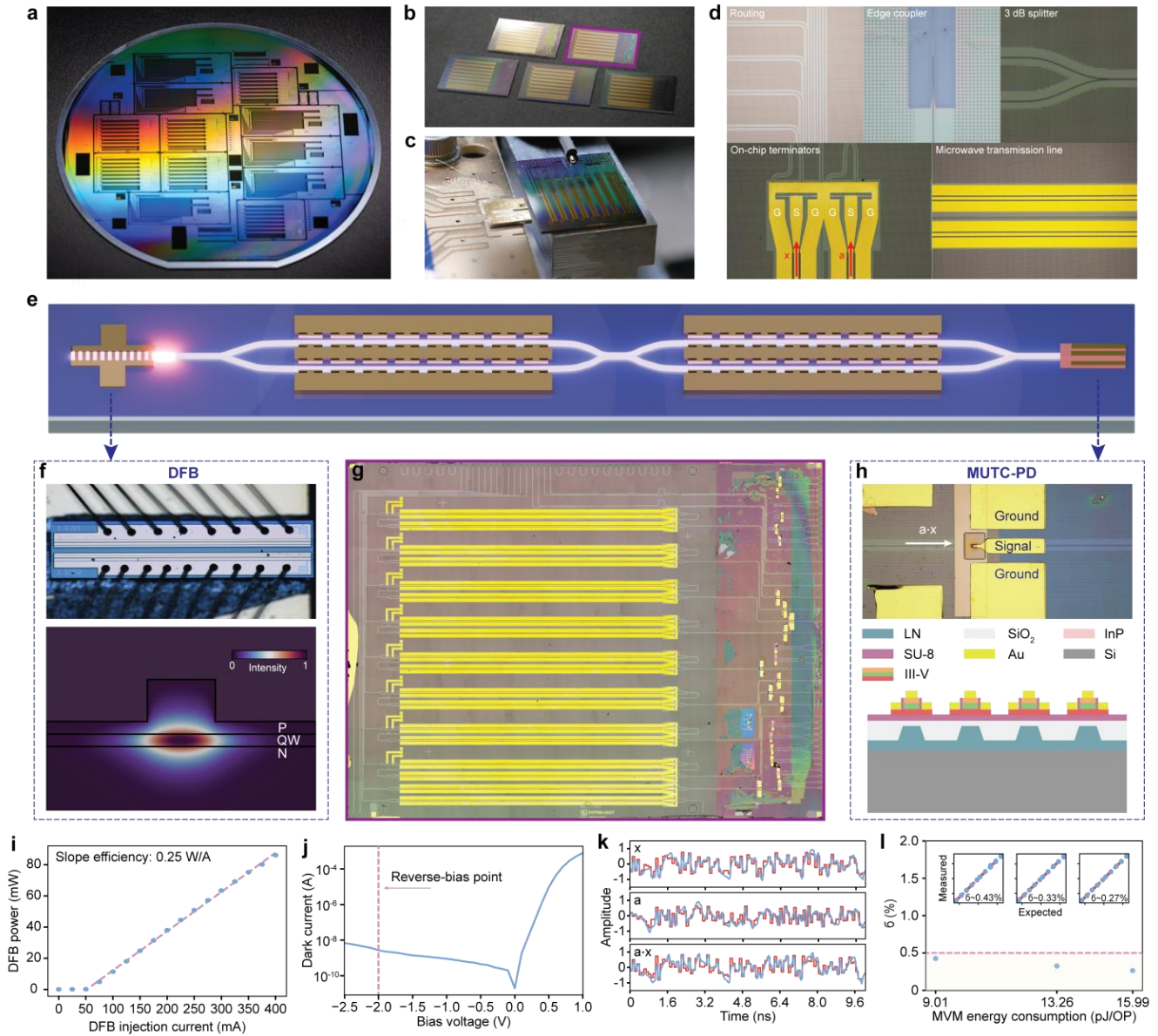

**Fig. 6 | Hybrid- and Heterogenous- integrated TFLN photonic computing circuit. a,** Wafer-scale fabrication of computing cores comprising a TFLN photonic computing circuit. **b,** Chiplets of TFLN computing cores from the wafer-scale process. **c,** Measurement setup for characterizing the hybrid- and heterogeneous-integrated system: light from a hybrid-integrated DFB laser source is butt-coupled to the TFLN computing core, while a heterogeneous-integrated MUTC-PD is used to perform optical-to-electronic conversion of the computing signal (the electrical signal is extracted by contact probes). **d,** Optical microscope images of essential building blocks used for the integrated TFLN circuit similar to Fig. 2**a**, except bilayer-taper edge couplers (top middle) are employed as a low-loss interface between the TFLN waveguide mode and the DFB waveguide mode. **e,** Schematic of a single channel integrated system. **f,** Optical microscope image of DFB laser and 2-D simulation of the DFB waveguide mode. P (N): positively- (negatively-) doped region; QW: quantum well. **g,** Optical microscope image of TFLN photonic computing circuit, with an array of cascaded amplitude modulators (left) and an array of MUTC-PDs (right). **h,** Optical microscope image of MUTC-PD and schematic of its cross section. **i,** DFB output power measured by an integrating sphere vs. injection current, with about 50 mA current threshold and 0.25 W/A slope efficiency. **j,** MUTC-PD dark current vs. bias voltage. For high-speed operation, a reverse-bias of -2 V is held for all measurements. **k,** Example waveforms of a temporally multiplexed computing operation between two random vectors $\vec{x}$ and $\vec{a}$ with 96.8 ps/symbol (10.33 GOPS per

channel). **l.** Computational accuracy $\sigma$, as previously defined, vs. computing energy consumption as the DFB output power is varied. Three energy consumptions are evaluated based on experimental conditions provided DFB injection currents of 85, 125, and 150 mA, respectively.